%% file: main.tex
\documentclass[letterpaper,twocolumn,10pt]{article}
\usepackage{usenix}

\usepackage{amsmath} 
\usepackage{amssymb}
\usepackage{graphicx}
\usepackage{tikz} 
\usetikzlibrary{tikzmark,calc}

\usepackage{booktabs}
\usepackage{threeparttable}
\usepackage{multirow}

\usepackage{pdfpages}
\usepackage{soul}
\usepackage{tcolorbox}
\tcbuselibrary{breakable}
\usepackage{enumitem}
\usepackage{listings}
\usepackage{microtype}

\begin{document}

\date{}

\title{\Large \bf TraceVIC: Causal Reasoning over Code Evolution for Identifying Vulnerability-Inducing Commits  \thanks{DISTRIBUTION STATEMENT A. Approved for public release. Distribution is unlimited. This material is based upon work supported by the Under Secretary of War for Research and Engineering under Air Force Contract No. FA8702-15-D-0001 or FA8702-25-D-B002. Any opinions, findings, conclusions or recommendations expressed in this material are those of the author(s) and do not necessarily reflect the views of the Under Secretary of War for Research and Engineering.}}  

\author{
{\rm Fnu Tanish}\\
Northern Illinois University \\
ttanish1@niu.edu
\and
{\rm Samiha Shimmi}\\
Northern Illinois University \\
sshimmi@niu.edu
\and
{\rm Samikshya Chapagain}\\
Northern Illinois University \\
schapagain2@niu.edu
\and
{\rm Hamed Okhravi}\\
MIT Lincoln Laboratory \\
hamed.okhravi@ll.mit.edu
\and
{\rm Mona Rahimi}\\
Northern Illinois University \\
rahimi@cs.niu.edu
\and
{\rm Lei Zhang}\\
Northern Illinois University \\
zhanglei@niu.edu
}

\maketitle

\begin{abstract}
Software vulnerabilities are often discovered long after they are introduced, making it difficult to identify the vulnerability-inducing commit (VIC) responsible for introducing the underlying vulnerable condition. Existing VIC identification techniques largely rely on \texttt{git blame} to trace vulnerable code through revision history and use positional heuristics, such as selecting its earliest or most recent modification. However, the true VIC may occur anywhere within this history, and vulnerable behavior may depend on code that evolves across multiple revisions. We therefore argue that VIC identification requires reasoning about \textit{how} vulnerability-relevant code evolves, rather than simply \textit{where} a candidate commit appears in the revision history.  

We present \textit{TraceVIC}, a temporal graph-based approach for identifying and ranking VICs by reasoning over code evolution. TraceVIC first localizes likely root-cause lines and traces their histories across revisions, constructing graph representations that capture program structure within each revision and the evolution of vulnerability-relevant code across the history. It reasons over the resulting revision history, using temporal edges to preserve correspondences between program elements across consecutive revisions, and directly ranks candidate commits according to their contribution to the vulnerable condition.  

Ablation results show that modeling the full revision history improves F2 from 0.637 to 0.814. TraceVIC improves F2 by up to 28.7\% over state-of-the-art methods and identifies a valid VIC for 78 of 79 vulnerabilities across four unseen C/C++ projects.
\end{abstract}


\section{Introduction}
Identifying Vulnerability-Inducing Commits (VICs) is fundamental to understanding how software vulnerabilities originate and supporting downstream tasks such as vulnerability remediation, affected-version identification, and learning-based vulnerability detection. Yet, vulnerabilities are rarely discovered when introduced~\cite{di2008evolution, pham2010detection, kim2006automatic}; empirical studies show that they persist in codebases for an average of 1,732 days before being patched~\cite{alexopoulos2022long}. Consequently, the commit that originally introduced the vulnerable condition---the VIC---often remains unknown~\cite{guo2026accurate, nguyen2025toward}. Incorrectly attributing the origin of a vulnerability can lead to incomplete fixes~\cite{liu2025characteristics, li2023commit}, inaccurate affected-version information in vulnerability databases such as NVD~\cite{nvd}, and noisy labels in datasets used to train vulnerability detection and mining models~\cite{nguyen2013unreliability, croft2022noisy}.

Prior work identifies VICs by tracing vulnerability-relevant code backward through revision history, typically using \texttt{git blame} or similar line-level history analysis~\cite{williams2008szz, borg2019szz}. The most widely used approach, SZZ~\cite{sliwerski2005changes}, starts from lines deleted by a vulnerability-fixing commit (VFC) and attributes them to the \textit{most recent} commits that previously modified those lines. However, SZZ was originally developed for general defects rather than security vulnerabilities. Unlike general defects, vulnerabilities can remain latent for extended periods; empirical studies report that more than 50\% of them are foundational, persisting across multiple software versions before they are discovered~\cite{ozment2006milk, pham2010detection}. Consequently, in the context of code vulnerabilities, the most recent modification of vulnerable code does not necessarily correspond to the commit that introduced the vulnerable condition.

Variants of SZZ such as V-SZZ~\cite{bao2022v} address this limitation by tracing vulnerable lines backward and selecting their \textit{earliest} modification. Yet both strategies rely on a positional assumption: the VIC is determined by \textit{where} a candidate appears in the revision history---for example, the latest or earliest modification---rather than by whether its change actually introduced the vulnerable condition~\cite{neto2018impact, borg2019szz, bosu2014identifying}. This assumption becomes particularly problematic when vulnerable behavior emerges as code evolves across multiple revisions. In such cases, the true VIC may be the earliest, latest, or an intermediate modification, and there is no principled way to determine the correct position a priori. Positional heuristics can therefore miss the true origin of vulnerabilities whose relevant code evolves through multiple revisions before fix~\cite{rosa2023comprehensive, lyu2024evaluating, herbold2022problems}.

More recent learning-based SZZ variants improve the analysis of individual code changes but do not fully overcome this limitation. For example, NeuralSZZ~\cite{tang2023neural} uses graph neural networks to model relationships among changes within a fixing commit, yet its analysis remains confined to a single revision~\cite{le2024latent, croft2022noisy}. 

Similarly, LLM-SZZ~\cite{fan2025llm} uses large language models to improve root-cause line selection while tracing revisions backward, but ultimately relies on V-SZZ's positional stopping criterion to identify the inducing commit. 
Thus, despite increasingly sophisticated analysis, existing approaches do not explicitly reason about \textbf{how} the vulnerable condition develops across the sequence of code revisions~\cite{chen2025vav, rosa2023comprehensive}.

In practice, vulnerabilities arise within evolving program states~\cite{yang2025code, bosu2014identifying}. A change may introduce a vulnerable condition, subsequent revisions may preserve, transform, or propagate it, and later changes may expose or eventually repair it~\cite{ozment2006milk, zimmermann2007predicting, perl2015vccfinder, li2018vuldeepecker}. Identifying the VIC therefore requires reasoning not only about individual revisions, but also about how vulnerability-relevant code evolves across the revision history.

\textbf{In this paper, we formulate VIC identification as reasoning over evolving program states.} Rather than identifying a commit based on its position in the revision history, our goal is to determine which change most strongly contributed to the emergence of the vulnerable condition. This requires reasoning about \textbf{what} changes across revisions and \textbf{how} vulnerability-relevant code persists or transforms throughout its evolution history. We refer to this formulation as \emph{contribution-based identification over code evolution}: candidate commits are evaluated according to their contribution to the vulnerable condition rather than their temporal position.

\textbf{We operationalize this formulation through \textit{TraceVIC}, a
temporal graph-based framework for identifying and ranking VICs.}
TraceVIC models vulnerability-relevant code across its revision history
rather than from a single program snapshot. Starting from the fixing
commit, it traces relevant code through prior revisions and constructs
a sequence of graph representations that capture the structural and
semantic relationships within each revision. TraceVIC then connects
corresponding code elements across revisions through temporal edges, representing how the code evolves over time. A graph-based
architecture reasons over this evolution to first rank likely root-cause
lines and then rank candidate commits according to their contribution to
the vulnerability. This enables TraceVIC to identify the VIC
without assuming that it occurs at a predefined position in the revision
history.

This formulation introduces two challenges: capturing vulnerability-relevant information across multiple revisions while preserving both within-revision program structure and cross-revision evolution, and distinguishing changes that contribute to the vulnerable condition from those that merely precede or follow it. TraceVIC addresses these challenges by jointly modeling revision-level program structure, cross-revision evolution, and commit-level relationships.

We investigate the following research questions:
\begin{itemize}
    \item \textbf{RQ1:} How does reasoning over code evolution improve VIC identification, and which components contribute to its effectiveness?
    \item \textbf{RQ2:} How well does TraceVIC generalize across projects?
\end{itemize}

An ablation study confirms that the primary benefit comes from reasoning over the full revision history, increasing F2 from 0.637 to 0.814 compared with single-revision analysis; explicit cross-revision correspondences provide an additional recall-oriented improvement.

We evaluate TraceVIC on manually validated vulnerabilities from the Linux kernel against state-of-the-art retrieval-, selection-, and ranking-based VIC identification methods. At top-3, TraceVIC improves F2 by 23.1\%, 17.3\%, and 28.7\% over the strongest retrieval-, selection-, and ranking-based baselines, respectively. 

We further evaluate TraceVIC on 79 vulnerabilities from four unseen C/C++ projects---FFmpeg, ImageMagick, OpenSSL, and PHP-SRC. Without training on these projects, TraceVIC identifies at least one valid VIC for \textbf{78 of 79 vulnerabilities} and achieves an overall precision of \textbf{0.855}, recall of \textbf{0.839}, F1 of \textbf{0.847}, and F2 of \textbf{0.842}. Compared with the strongest baseline on this dataset (F1=0.700), TraceVIC improves F1 by \textbf{21.0\%}. These results demonstrate that reasoning over temporal code evolution provides consistent improvements over existing VIC identification formulations and generalizes beyond the project used for training.

In summary, this paper makes the following contributions: 
\begin{itemize}[leftmargin=*]  
\item \textbf{Evolution-Aware VIC Formulation.} We formulate VIC identification as contribution-based reasoning over evolving program states, moving beyond approaches that attribute vulnerabilities using fixed positional heuristics.  
\item \textbf{Temporal Graph-Based VIC Identification.} We introduce \textit{TraceVIC}, a framework that represents vulnerability-relevant code across multiple revisions and augments revision-level program graphs with explicit correspondences between program elements across consecutive revisions. TraceVIC uses these representations to localize root-cause lines and directly rank candidate VICs.  
\item \textbf{Comprehensive Empirical Evaluation.} We evaluate TraceVIC against state-of-the-art retrieval-, selection-, and ranking-based VIC identification methods, isolate the contributions of temporal modeling through ablations, and evaluate generalization across four unseen C/C++ projects.
\end{itemize}

In this paper, we focus on vulnerabilities in C/C++ systems, where low-level memory and pointer operations make accurate vulnerability attribution particularly challenging. All code, trained models, datasets, and scripts required to reproduce our results are publicly available in our repository. \footnote{public repository: \url{https://anonymous.4open.science/r/TraceVIC/}}.

\section{A Motivating Example for CVE-2014-2309: Why Positional Reasoning Fails?}
Existing SZZ-based approaches trace vulnerability-relevant code through revision history and identify VICs using positional heuristics, typically selecting the most recent (B-SZZ) or earliest (V-SZZ) modification~\cite{sliwerski2005changes, bao2022v, rosa2023comprehensive}. However, vulnerable code may evolve across multiple revisions, and the true VIC can occur anywhere in this history~\cite{pianco2016code,alohaly2017changes}.


\begin{figure}[htbp]
    \centering
    \includegraphics[
    width=\columnwidth,
]{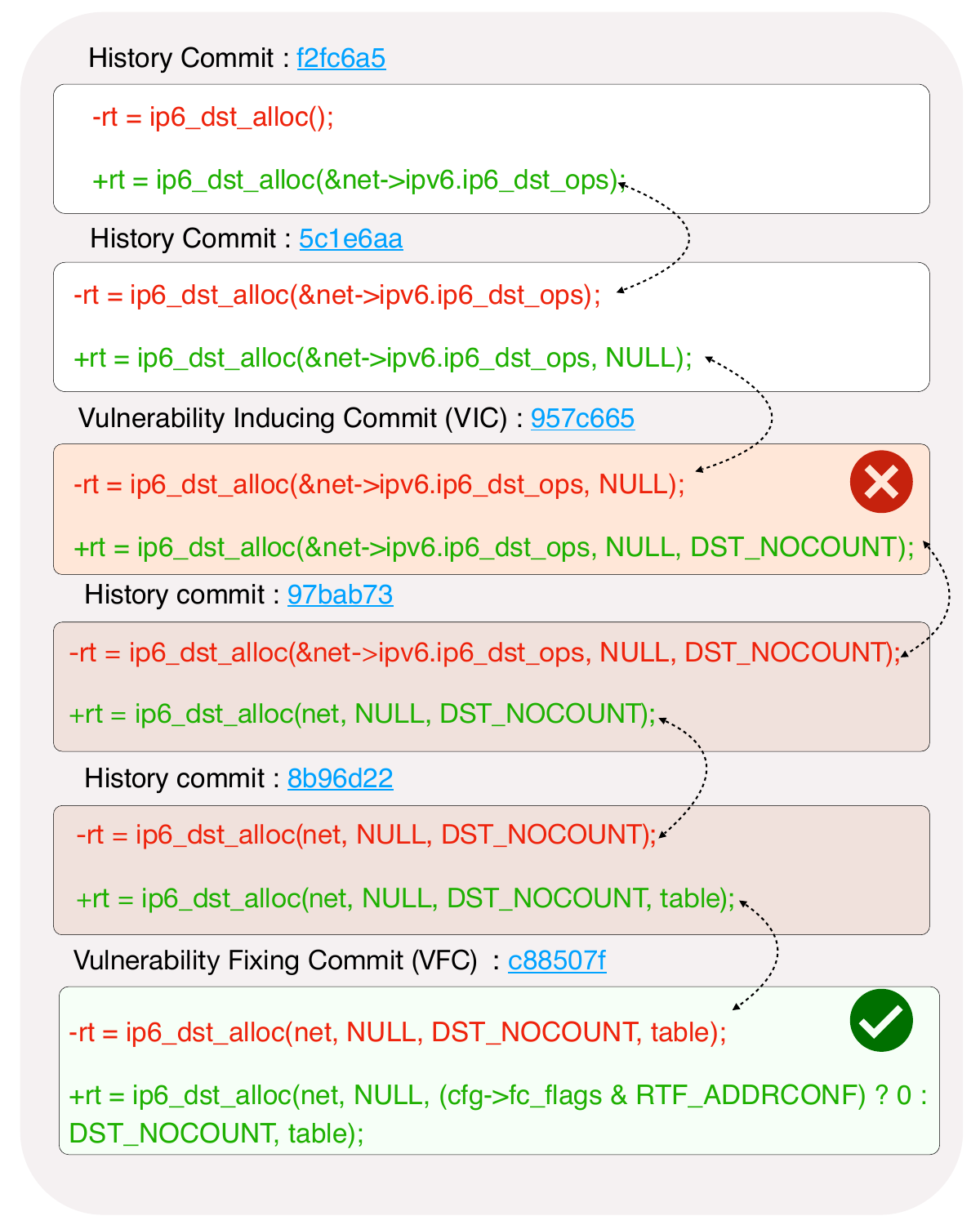}

    \caption{Vulnerability-Inducing Commit occurring at a middle position  (CVE-2014-2309) in the history chain.
    }
    \label{fig:middle_commit}
\end{figure}

\begin{figure*}[t]
    \centering
    
    \includegraphics[page=1, 
        width=0.89\textwidth, 
        keepaspectratio]{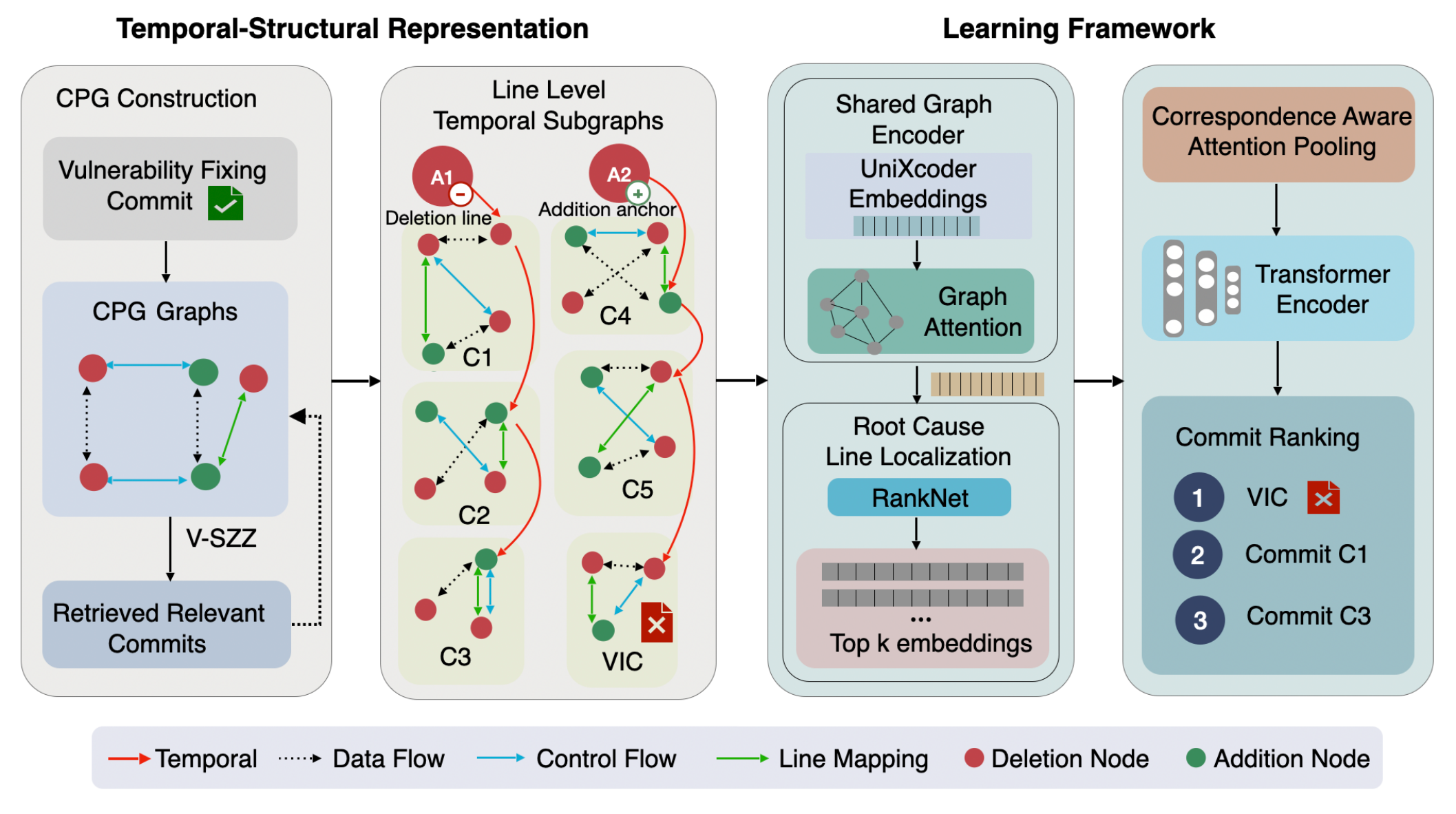}
   
    \caption{Overview of TraceVIC. TraceVIC constructs temporal Code Property Graphs (CPGs) over candidate commit histories, encodes structural and evolutionary relationships to localize root-cause lines, and aggregates the selected representations to rank candidate VICs.
    }
    \label{fig:pipeline}
\end{figure*}


We illustrate this limitation using CVE-2014-2309, a real world Linux kernel vulnerability that allows a remote attacker to exhaust system memory through unbounded allocation of routing entries. Figure~\ref{fig:middle_commit} shows the evolution of the relevant \texttt{ip6\_dst\_alloc} call across several revisions.

Early commits establish and extend the basic allocation logic without introducing unsafe behavior. A later commit, \texttt{957c665}, introduces the semantic change that enables route entries to bypass the relevant accounting mechanism, creating the vulnerable condition. Subsequent commits preserve and further evolve this behavior before the vulnerability is eventually fixed. Thus, the ground-truth VIC occurs in the middle of the revision history rather than at either boundary.

A positional method can therefore select the wrong commit even when it correctly traces the relevant code history. B-SZZ favors a later modification, while V-SZZ favors an earlier one; neither evaluates which candidate change most strongly corresponds to the introduction of the vulnerable condition. This motivates formulating VIC identification as a ranking problem over the candidate history: rather than treating all retrieved commits equally or selecting a candidate based on temporal position, the model prioritizes the ground-truth VIC. In this example, the ranking places 957c665 above the other commits in the history because it is the commit that introduced the vulnerable condition.

This example motivates treating VIC identification as attribution over code evolution rather than positional selection. A method must therefore reason over the sequence of vulnerability-relevant program states and identify which change introduced the vulnerable condition.

\section{Problem Formulation}
\label{sec:formulation}

This section formalizes the problem of VIC identification:\\
 



\noindent\textbf{Vulnerability-fixing commit (VFC):}
A fixing commit $V$ modifies one or more source files to resolve a reported security vulnerability. Its diff contains deleted lines
$\mathcal{D}(V)=\{d_1,\ldots,d_m\}$ and added lines
$\mathcal{A}(V)=\{a_1,\ldots,a_n\}$, where $m+n\geq1$. We distinguish two types of fixes:
\begin{itemize}[nosep,leftmargin=*]
\item \emph{Deletion-involving} ($m\geq1$): the diff removes at least one existing line, including pure deletions, modifications/replacements, or relocations. TraceVIC initiates history tracing from the deleted lines.
\item \emph{Addition-only} ($m=0,\;n\geq1$): the diff adds lines without deleting existing ones, leaving no deleted line to trace. TraceVIC uses nearby pre-existing code as anchors.\\
\end{itemize}

\noindent\textbf{Anchor-Line Extraction for Addition-Only Fixes:}
\label{sec:anchor}
Most SZZ-family methods, except TSE-SZZ, construct VIC candidates by tracing only lines deleted by a fixing commit, based on the assumption that the vulnerability originated only from code removed by the fix. However, some vulnerabilities are repaired entirely by adding previously missing code, leaving no deleted line to trace. In our dataset, 183 of 755 cases (24.2\%) are such addition-only fixes, for which $\mathcal{D}(V)=\emptyset$. Conventional deletion-based tracing therefore produces no VIC candidates for these cases.

TraceVIC handles these cases through \emph{anchor-line extraction}. Although the lines added by the fix did not exist in the pre-fix revision and therefore cannot be traced backward, the existing lines surrounding the insertion point can be traced through revision history. TraceVIC therefore uses pre-existing code around the insertion point as anchors for locating the vulnerability-relevant revision history. These anchors are used only to initiate history tracing; the commits to which they trace are subsequently evaluated and ranked by TraceVIC rather than being directly treated as VICs.

For each contiguous block of added lines, TraceVIC extracts up to three anchors from the pre-fix revision: the nearest meaningful lines above and below the insertion and the enclosing function signature. Structural boilerplate and simple error-handling statements are filtered to avoid uninformative anchors. Each anchor independently initializes the same history-tracing procedure used for deleted lines, and the resulting candidate histories are subsequently evaluated.

Multiple anchor types are extracted because no single neighbouring line is reliable
across fixes. Which anchor reaches the inducing commit depends on the structure of the
defect: for a missing guard, the operation requiring protection may sit either above
or below the inserted check, and only one of the two may carry the relevant history;
for a missing cleanup, the acquisition of the leaked resource typically precedes the
insertion. Extracting complementary anchors and filtering uninformative ones therefore
recovers the vulnerability-relevant history in cases where any individual anchor would
fail. Appendix~\ref{app:addition_cases_structural} details the extraction procedure and traces three
addition-only CVE fixes through it, including one in which the resulting attribution
independently matches the kernel maintainers' own \texttt{Fixes:} tag.

Throughout the remainder of the paper, we use $\mathcal{L}(V)$ to denote the set of lines used to initiate history tracing:
\begin{equation}
\mathcal{L}(V)=
\begin{cases}
\mathcal{D}(V), & \mathcal{D}(V)\neq\emptyset,\\
\mathcal {A}(V)={H}(V), & \mathcal{D}(V)=\emptyset,
\end{cases}
\end{equation}
where $\mathcal{H}(V)$ denotes the extracted anchor set. Each
$l_i \in \mathcal{L}(V)$ is processed through the same candidate-chain
and temporal-graph construction pipeline. Additional details and examples
of anchor extraction are provided in
Appendix~\ref{app:addition_cases_structural}.\\

\noindent\textbf{Candidate Commit Chain:} For each line $l_i \in \mathcal{L}(V)$, we trace its modification history backward through the repository, yielding a \emph{candidate commit chain}:
\begin{equation}
  \mathcal{C}^{(i)} = \langle c^{(i)}_1,\; c^{(i)}_2,\; \ldots,\; c^{(i)}_{n_i} \rangle
\end{equation}
where $c^{(i)}_1$ is the most recent commit that modified $l_i$ prior to the fix, and $c^{(i)}_{n_i}$ is the earliest traceable modification. Since a single commit may modify multiple lines, it can appear in more than 
one candidate chain. To preserve the per-chain context of each occurrence, the overall 
\emph{candidate set} is defined as the multiset union of all chains:
\begin{equation}
    \mathcal{C}(V) = \biguplus_{i=1}^{m} \mathcal{C}^{(i)},
\end{equation}
where $\biguplus$ denotes multiset union and m is number of traced lines. Each element in $\mathcal{C}(V)$ is thus a tuple $(c, i)$, representing commit $c$ in the context of chain $\mathcal{C}^{(i)}$.

For addition-only commits, each anchor line is emitted as a synthetic deletion line, so the chain construction above applies uniformly.\\

\noindent\textbf{Vulnerability-inducing Commit (VIC):} We define the VIC as the commit that introduces or modifies the program semantics that create the vulnerable condition addressed by the VFC. A later commit that merely exposes an existing vulnerability, or an earlier commit that introduces code subsequently involved in the vulnerability without creating the vulnerable condition, is therefore not considered the VIC. A VIC must therefore contributed in creating the vulnerability; merely modifying code that is later involved in the vulnerability is insufficient. We follow the expert-validated ground truth in the benchmark dataset.\\
 
\noindent\textbf{VIC Identification:} We formulate VIC identification problem as a \textbf{commit ranking problem}:

\textbf{Given:} a vulnerability-fixing commit $V$, its traced lines $\mathcal{L}(V) = \{l_1, \dots, l_m\}$, and the corresponding candidate commit chains $\{\mathcal{C}^{(i)}\}_{i=1}^{m}$
\textbf{Goal:} learn a scoring function $f(c, i) = f\!\left(c \mid \mathcal{C}^{(i)}\right)$ 
over each tuple $(c, i) \in \mathcal{C}(V)$. The final score for a commit $c$ is 
aggregated across all chains in which it appears:
\begin{equation}
    \hat{f}(c) = \max_{i:\, c \in \mathcal{C}^{(i)}} f\!\left(c \mid \mathcal{C}^{(i)}\right)
\end{equation}
such that $\hat{f}(c^{*}) > \hat{f}(c)$ for all $c \in \mathcal{C}(V) \setminus \{c^{*}\}$, 
where $c^{*}$ is the true VIC, identified as 
$c^* = \arg\max_{c \in \mathcal{C}(V)} \hat{f}(c)$.

The key distinction between TraceVIC and prior work lies in how this function $f$ is defined. Existing SZZ-based methods implicitly reduce $f$ to a deterministic positional rule:
\begin{equation}
    c^*_{\text{B-SZZ}} = c_1, \qquad c^*_{\text{V-SZZ}} = c_n
\end{equation}
where $c_1$ is the most recent and $c_n$ the earliest commit in a candidate chain. In both cases, the inducing commit is selected based solely on its position, without considering the semantic content of changes.

Unlike positional approaches, TraceVIC learns $f$ over temporal-structural representations of candidate commits and identifies the VIC as $\arg\max_{c \in \mathcal{C}(V)} f(c)$. These representations capture both the properties of vulnerability-relevant code within each revision and its evolution across revisions, allowing the ground-truth VIC ($c^*$) to receive the highest score regardless of its position in the candidate history. Although TraceVIC uses positional encodings to preserve the ordering of revisions, its ranking objective does not impose a predefined positional rule, such as selecting the earliest or most recent modification. Thus, temporal position provides evolutionary context rather than determining the VIC directly.

This ranking formulation differs from prior learning-based approaches
in both the information considered and how the final VIC is identified.
NeuralSZZ analyzes code changes within individual revisions, whereas
LLM-SZZ improves semantic reasoning over vulnerability-relevant lines
but ultimately relies on V-SZZ's positional stopping criterion. In
contrast, TraceVIC jointly reasons over the evolution of code across
multiple revisions and directly ranks candidate commits without assuming
that the VIC occurs at a predefined temporal position.

TraceVIC operationalizes our formulation through two key components:
(i) a temporal-structural representation that captures vulnerability-relevant
code and its evolution across revisions, and (ii) a ranking objective that
evaluates candidate commits using this representation.

\section{Temporal-Structural Representation}
\label{sec:overview}

Given a
vulnerability-fixing commit $V$, TraceVIC produces a ranked list of
candidate commits, with the highest-scoring candidate identified as the
predicted VIC. As shown in Figure~\ref{fig:pipeline}, the framework
consists of \textit{temporal-structural graph construction} followed by two
sequential \textit{learning stages}.This section describes the graph construction.

\subsection{Revision-Level Structure}
\label{sec:structural}

For each traced line $l_i \in \mathcal{L}(V)$, TraceVIC retrieves its
modification history as a candidate chain
$\mathcal{C}^{(i)}=\{c_1,\ldots,c_n\}$. TraceVIC identifies the source
file containing $l_i$ in the fixing commit and follows that file across
the candidate chain. For each commit $c_t \in \mathcal{C}^{(i)}$, it
constructs a revision-level Code Property Graph (CPG)
$\mathcal{G}_t=(V_t,E_t)$ for that file and then restricts the graph to
the code changed by $c_t$. 

Specifically, TraceVIC represents program structure using AST-level
units, including functions, control structures, statements, and
expressions, as graph nodes. Accurate AST construction for C/C++ requires
the compilation context, including headers, macros, and compiler options.
TraceVIC therefore reconstructs this context from the project build
before parsing the source code. In our implementation, Bear~\cite{Bear}
captures the compilation context, which Clang~\cite{lattner2004llvm}
uses to construct the AST.

To capture relationships among these program elements, TraceVIC augments
the AST nodes with control- and data-flow dependencies. We use
Joern~\cite{yamaguchi2014modeling} to extract CFG and DFG edges and align
them with the AST nodes at the source-line level. After constructing the
CPG, TraceVIC retains the AST nodes whose source ranges overlap lines
added or deleted by $c_t$ in the traced file, together with the
relationships among those nodes. The resulting $\mathcal{G}_t$ therefore
captures the structural and flow relationships among the code elements
changed by $c_t$ in the file associated with $l_i$.

\subsection{Cross-Revision Node Correspondence}
\label{sec:temporal_edges}

After constructing the revision-level CPGs, TraceVIC identifies the CPG
node corresponding to the traced line in each revision and connects
these nodes across consecutive revisions. This provides an explicit
representation of how the traced code persists or changes throughout
its modification history.

For each pair of consecutive commits $(c_t,c_{t+1})$, TraceVIC identifies
the CPG node representing the traced code in each revision. Because a
CPG node may span multiple source lines, and the traced code may change
or shift to a different line number across revisions, TraceVIC uses a
three-level matching strategy. It first matches using both source-line
range and code prefix. If no match is found, it uses source-line range
alone to accommodate changes in code content, and finally code prefix
alone to accommodate line-number shifts.

Once the corresponding nodes are identified, TraceVIC connects them
with bidirectional temporal edges, allowing information to propagate
across revisions in both directions. If the CPG exists but no
corresponding node can be identified, no temporal edge is added. If a
revision produces an empty CPG, TraceVIC instead creates a synthetic
node from the traced history entry to preserve that revision in the
evolution sequence. Appendix~\ref{app:node_correspondence} provides the
complete matching procedure.

\subsection{Temporal Subgraph Formation}
\label{sec:subgraph}

For each traced line $l_i \in \mathcal{L}(V)$, TraceVIC has a candidate
chain $\mathcal{C}^{(i)}=\{c_1,\ldots,c_n\}$ and a corresponding
revision-level CPG $\mathcal{G}_t=(V_t,E_t)$ for each commit $c_t$.
TraceVIC combines these CPGs into a single temporal subgraph
$\mathcal{G}^{(i)}$ by retaining their intra-revision edges and adding
the temporal edges established between consecutive revisions:
\begin{equation}
    \mathcal{G}^{(i)} =
    \left(
    \bigcup_{t=1}^{n} V_t,\quad
    \bigcup_{t=1}^{n} E_t \;\cup\; E_{\text{temp}}
    \right).
\end{equation}

Within each revision,
\begin{equation}
    E_t =
    E^{\text{cfg}}_t
    \cup E^{\text{dfg}}_t
    \cup E^{\text{lm}}_t,
\end{equation}
where $E^{\text{cfg}}_t$ and $E^{\text{dfg}}_t$ capture control- and
data-flow dependencies, respectively, and $E^{\text{lm}}_t$ captures
the local structural relationships between AST-level nodes. Across
revisions,
\begin{equation}
E_{\text{temp}}
=
\left\{
(v_t,v_{t+1}),\,(v_{t+1},v_t)
\;\middle|\;
v_t \leftrightarrow v_{t+1},\
t=1,\ldots,n-1
\right\},
\end{equation}
where $v_t \leftrightarrow v_{t+1}$ indicates that the two nodes were
identified as corresponding program elements using the matching
procedure in Section~\ref{sec:temporal_edges}.

The resulting $\mathcal{G}^{(i)}$ therefore represents the complete
history of $l_i$ in a single graph: CFG, DFG, and LINEMAP edges capture
relationships within each revision, while temporal edges connect the
corresponding code across revisions. These relationships are represented
by seven edge types: CFG\_FWD/\allowbreak BWD, DFG\_FWD/\allowbreak BWD,
LINEMAP, and TEMPORAL\_FWD/\allowbreak BWD. The graph encoder learns a
separate bias for each edge type so that structural and temporal
relationships remain distinguishable during message passing.

TraceVIC constructs a separate temporal subgraph $\mathcal{G}^{(i)}$
for each $l_i \in \mathcal{L}(V)$ because different traced lines may
follow different modification histories. Within each subgraph, the CPG
node corresponding to $l_i$ is assigned index~0, providing a consistent
location from which TraceVIC extracts its representation for the
subsequent line-level scoring stage.

\section{Learning Framework}
\label{sec:learning}
\label{sec:encoder}
 
TraceVIC identifies the VIC through two sequential learning stages that share a common graph encoder. The encoder learns representations of the temporal subgraphs constructed above; it is trained jointly with the first-stage line-ranking task and then frozen for the second-stage commit-ranking task.

 \subsection{Shared Graph Encoder}
For each temporal subgraph $\mathcal{G}^{(i)}$, TraceVIC initializes
each node $v$ with a UniXcoder~\cite{guo2022unixcoder} embedding
$\mathbf{x}_v \in \mathbb{R}^{d_u}$ of the code element represented by
that node. The embedding is projected to the encoder dimension $d$ and
combined with a fixed sinusoidal positional encoding (PE) of the same
dimension:
\begin{equation}
    \mathbf{h}_v =
  Proj(\mathbf{x}_v) +
    \operatorname{PE}(\operatorname{pos}(v)),
    \qquad
    \mathbf{h}_v \in \mathbb{R}^{d},
\end{equation}
where ${h}_v $ is the initial representation of node $v$ and
$\operatorname{pos}(v)\in\{0,\ldots,n-1\}$ denotes the position of the
commit containing $v$ in the candidate chain. The positional encoding preserves the ordering
of revisions as contextual information; it does not impose a rule that
the VIC must occur at any particular position.

The initialized node representations are then propagated over the
temporal subgraph using edge-aware multi-head graph attention.
For each graph edge, the attention score includes a learned bias
specific to its edge type and attention head. This allows the encoder
to distinguish information propagated through control-flow, data-flow,
local structural, and temporal relationships rather than treating all
graph connections equivalently. The encoder therefore jointly
represents the code associated with each node, its position in the
revision sequence, and its structural and temporal relationships with
other nodes. Full encoder equations are provided in
Appendix~\ref{app:encoder_equations}.
 
\subsection{Relevant Line Localization}
\label{sec:localization}

The first learning stage ranks the traced lines
$l_i \in \mathcal{L}(V)$ to identify those whose modification histories
are most likely to be relevant to the vulnerability.

\textbf{Label derivation:}
Because ground truth is available at the commit level, TraceVIC derives
line-level labels from each traced line's candidate chain. A traced line
$l_i$ receives a positive label if its candidate chain
$\mathcal{C}^{(i)}$ contains the ground-truth VIC $c^*$:
\begin{equation}
    y_i = \mathbf{1}\bigl[c^* \in \mathcal{C}^{(i)}\bigr].
\end{equation}
Thus, if multiple traced lines lead to candidate chains containing
$c^*$, each receives $y_i=1$, which means that tracing $l_i$
 produces a chain containing the ground-truth VIC.

\textbf{Line ranking:}
After encoding each temporal subgraph $\mathcal{G}^{(i)}$, TraceVIC
extracts the final representation of its traced-line node (node~0):
\begin{equation}
    \mathbf{e}^{(i)} = \mathbf{h}_{0}^{(L)} \in \mathbb{R}^{d},
\end{equation}
where $\mathbf{h}_{0}^{(L)}$ is the representation of node~0 after
$L$ graph-attention layers, and $\mathbf{e}^{(i)}$ denotes the resulting
representation for traced line $l_i$. At this point,
$\mathbf{e}^{(i)}$ captures the code semantics, structural dependencies within revisions, and cross-revision evolution associated with $l_i$.

A RankNet~\cite{burges2010ranknet} scoring head assigns a score to each traced-line representation and is trained pairwise to give higher scores to lines whose candidate chains contain the ground-truth VIC than to lines whose chains do not. At inference, lines are ordered by these scores, and the top-$k$ lines are passed to the commit-attribution stage.

\subsection{Commit-Level Attribution}
\label{sec:attribution}

The second learning stage determines which candidate commit most strongly
contributed to the vulnerable condition. It takes the top-$k$ temporal
subgraphs selected by the line-localization stage and leverages their node
representations produced by the shared graph encoder, which is frozen
during this stage.
TraceVIC first aggregates all nodes
associated with each commit into a single commit representation. It then
models the resulting sequence of commit representations and assigns a
score to each candidate commit.

\textbf{Correspondence-aware attention pooling:}
For each candidate commit $c_t$, TraceVIC collects the encoded nodes
belonging to $c_t$ across the selected top-$k$ temporal subgraphs. These
nodes contain two types of information: (i) \emph{Correspondence nodes} are
connected to a corresponding code element in an adjacent revision by a
temporal edge and therefore carry information about code evolution, and (ii)
\emph{Local nodes} have no such temporal correspondence and capture
structural information specific to that revision.

Rather than pooling both types identically, TraceVIC uses attention to
learn how strongly each node should contribute to the commit
representation. It maintains two learnable query vectors for each
attention head $h$: $q_{\mathrm{corr}}^h$ for nodes participating in
cross-revision correspondences and $q_{\mathrm{local}}^h$ for nodes
carrying only revision-local information. For a node $v$, the query is
selected as
\begin{equation}
    q_v^h =
    \begin{cases}
        q_{\mathrm{corr}}^h, & \text{if } v \text{ participates in a temporal correspondence},\\
        q_{\mathrm{local}}^h, & \text{otherwise}.
    \end{cases}
\end{equation}
where $h$ denotes the attention head. Each query determines the attention
assigned to its corresponding node type, and the weighted node
representations are aggregated into a single representation
$\mathbf{c}_t$ for commit $c_t$. This allows the model to preserve the
distinction between cross-revision evolutionary information and
revision-local structural information when representing a commit. Full
pooling equations are provided in Appendix~\ref{app:phase2_equations}.

\textbf{Commit-sequence modeling:}
The resulting commit representations
$\langle\mathbf{c}_1,\ldots,\mathbf{c}_n\rangle$ are processed by a
two-layer Transformer encoder~\cite{vaswani2017attention}. This allows
each candidate commit to be evaluated in the context of the complete
candidate history rather than independently.

\textbf{Commit ranking:}
A two-layer feed-forward head maps each contextualized commit
representation to a scalar score. The highest-scoring candidate is
returned as the predicted VIC. Training uses
\begin{equation}
    \mathcal{L}
    =
    \mathcal{L}_{\mathrm{focal}}
    + \lambda\mathcal{L}_{\mathrm{margin}},
\end{equation}
where $\mathcal{L}_{\mathrm{focal}}$ is focal cross-entropy
loss~\cite{lin2017focal} over label-smoothed targets to address class
imbalance, and $\mathcal{L}_{\mathrm{margin}}$ encourages the
ground-truth VIC to receive a score at least $\delta$ greater than the
other candidate commits. The complete loss formulation is provided in
Appendix~\ref{app:phase2_equations}.
 
\subsection{Training Protocol}
\label{sec:training}

The two learning stages are trained sequentially. During root-cause line localization, UniXcoder, the projection layer, and graph-attention layers are trained jointly using the pairwise ranking objective, with fixed temporal positional encodings incorporated into the node representations. The bottom eight layers of UniXcoder are
frozen to preserve its general-purpose code representations. After this
stage converges, the shared graph encoder is frozen, and the commit-level
attribution stage trains only the correspondence-aware pooling,
commit-sequence Transformer, and ranking head.

Training is supervised using manually verified ground-truth
VICs provided in~\cite{jiang2024understanding}, rather than VIC labels generated by
SZZ or other positional heuristics. Consequently, the ranking objective
does not impose an earliest- or most-recent-commit rule: the model learns
to rank the ground-truth VIC highest regardless of where it occurs in
the candidate history.

\section{Evaluation}
\label{sec:evaluation}
We evaluate TraceVIC with respect to our two RQs. 
\subsection{Dataset and Metrics}
\label{subsec:metrics}

\textbf{Dataset:}
We evaluate TraceVIC on the Linux kernel vulnerability dataset from Jiang et al.~\cite{jiang2024understanding}, which originally contains 1,349 manually verified VIC instances. The dataset consists of curated triplets of CVEs, VFCs, and their corresponding ground-truth VICs. Unlike SZZ-generated labels, these inducing commits were manually validated by the original authors through commit history analysis and developer evidence.

Following the filtering process described in Appendix~\ref{app:dataset_filtering} — including removing oversized files, assembly files, missing repository files, and cases where no  candidate commit could be constructed—the final evaluation set contains 780 vulnerability instances.


\textbf{Metrics:}
For commit ranking, we report Precision@$k$, Recall@$k$, F1@$k$, and F2@$k$ for $k \in \{1, 2, 3\}$, measuring how many ground truth VICs are found within the top-$k$ ranked commits:
\begin{equation}
\small
\begin{aligned}
P@k &= \frac{\sum_{i=1}^{N} |\hat{C}_i^k \cap GT_i|}{\sum_{i=1}^{N} |\hat{C}_i^k|}, \qquad R@k = \frac{\sum_{i=1}^{N} |\hat{C}_i^k \cap GT_i|}{\sum_{i=1}^{N} |GT_i|}, \\
F1@k &= 2 \cdot \frac{P@k \cdot R@k}{P@k + R@k}, \qquad F2@k =  \frac{5 \cdot P@k \cdot R@k}{4 \cdot P@k + R@k}
\end{aligned}
\end{equation}
where $\hat{C}_i^k$ is the set of distinct commits among the top-$k$ ranked predictions for case $i$, and $GT_i$ is the set of ground truth inducing commits for case $i$.  Full metrics definitions for the rootcause-line ranking stage are provided in Appendix~\ref{app:metrics}.

Across all evaluations, we report precision, recall, F1, and F2. While F1 provides a balanced measure of precision and recall, F2 places greater emphasis on recall, which is particularly important for VIC identification because failing to recover the true inducing commit is more consequential than returning additional candidates for manual inspection.

\subsection{Experimental Protocol}
\label{subsec:experiment protocol}

\textbf{Evaluation Setup:}
TraceVIC and NeuralSZZ are learning-based and therefore require training data. Both are evaluated under an identical 5-fold cross-validation procedure over 755 cases remaining after removing 25 duplicated VIC-VFC pairs from the 780-instance filtered dataset. Within each fold, the held-out fold serves as the test set, and the remaining data are split 80/20 into training and validation sets, yielding approximately 483 training cases, 121 validation cases, and 151 test cases per fold. Each case is held out in exactly one fold, so five test folds jointly partition the dataset, and every instance is evaluated exactly once as unseen test data. \\
\textbf{Pooled evaluation:} We do not average per-fold metrics. Instead, predictions from all five held-out folds are pooled into a single prediction set covering all 755 cases, and the metrics of Eq. 12 are computed once over this pool. \\
\textbf{Baseline evaluation:} The SZZ-family variants (B-SZZ, V-SZZ, AG-SZZ, MA-SZZ, TSE-SZZ, RA-SZZ, L-SZZ, R-SZZ) are deterministic and untrained; each is executed once over the same 755 cases, yielding predictions directly comparable to the pooled learning-based predictions. The LLM-based methods are stochastic: LLM-SZZ is evaluated with Mistral-7B-Instruct-v0.3 and Gemini 2.5 Pro, while LLM4SZZ and AgenticSZZ are evaluated with Gemini 2.5 pro. 

\subsection{Impact of Temporal Modeling (RQ1)}
\label{sec:rq1}
We compare TraceVIC against existing VIC identification methods that do not explicitly model the temporal evolution of code changes. Based on their output formulation, we group these methods into three categories: \emph{retrieval-based}, \emph{selection-based}, and \emph{ranking-based} methods. Because these methods produce fundamentally different outputs, we report the results for each category separately. Retrieval-based methods return an unordered set of candidate VICs, selection-based methods produce a single final VIC prediction, and ranking-based methods prioritize candidates according to their likelihood of being the true VIC. 

This separation is important because a direct comparison of raw metrics across the three categories would conflate different prediction tasks. Retrieval-based SZZ variants are evaluated on their ability to \emph{retrieve} the true VIC within a candidate set; selection-based approaches are evaluated on their ability to \emph{select} the correct VIC as a single prediction; and ranking-based approaches are additionally evaluated on their ability to \emph{prioritize} the true VIC near the top of the candidate list. For ranking-based methods, we additionally report performance at top-1, top-2, and top-3 to evaluate not only whether the true VIC is recovered, but also how highly it is prioritized.

\subsubsection{Retrieval-based Methods}

Retrieval-based methods return an unordered set of candidate VICs. A prediction is therefore successful when the ground-truth VIC is contained anywhere in the retrieved candidate set. We evaluate six widely used retrieval-based baselines: B-SZZ, V-SZZ, AG-SZZ, MA-SZZ, TSE-SZZ, and RA-SZZ. These methods primarily differ in how they trace vulnerable lines through the revision history and how they filter the resulting candidate commits. Table~\ref{tab:retrieval_methods} summarizes these approaches.

\begin{table*}[t]
\centering
\small
\caption{Overview of \textit{Retrieval-based methods} for VIC identification.}
\label{tab:retrieval_methods}
\begin{tabular}{p{1.2cm} p{7.2cm} p{3.0cm} p{3.8cm}}
\toprule
\textbf{Method} & \textbf{What it does} & \textbf{Main Distinction} & \textbf{Category} \\
\midrule

B-SZZ &
Uses blame/annotation to trace lines changed by the fixing commit to the most recent commit that modified each line. &
Last modification &
Baseline tracing heuristic \\

V-SZZ &
Repeatedly traces vulnerable lines backward and identifies the earliest commit that modified the vulnerable code, rather than the most recent one. &
Earliest modification &
Historical tracing heuristic \\

AG-SZZ &
Extends B-SZZ by filtering non-semantic or cosmetic changes, such as whitespace, comments, and formatting. &
Filters cosmetic changes &
Candidate-filtering heuristic \\

MA-SZZ &
Extends AG-SZZ by excluding meta-changes, such as branch, merge, and property changes that do not directly modify source behavior. &
Filters meta-changes &
Candidate-filtering heuristic \\

TSE-SZZ (VCC) &
Extends SZZ by additionally blaming contextual lines surrounding added-only code blocks, under the premise that such additions often represent missing validation checks. &
Blames contextual lines &
Baseline tracing heuristic \\

RA-SZZ &
Filters candidate commits associated with refactoring operations to avoid identifying behavior-preserving changes as vulnerability-inducing. &
Filters refactorings &
Candidate-filtering heuristic \\

\bottomrule
\end{tabular}
\end{table*}

For consistency, we use the implementations of B-SZZ, V-SZZ, AG-SZZ, TSE-SZZ, RA-SZZ and MA-SZZ provided by \cite{hinrichs2026back}. All implementations, configurations, and replication materials used in our evaluation are available in our public repository.

\begin{table}[htbp]
\small
    \centering
    \caption{VIC identification results comparing TraceVIC with \textit{retrieval-based} methods.}
    \label{tab:retrieval-results}
    \begin{tabular}{lcccc}
        \toprule
        \textbf{Method} & \textbf{Precision} & \textbf{Recall} & \textbf{F1} & \textbf{F2} \\
        \midrule
        B-SZZ   & 0.498 & 0.693 & 0.580 & 0.643 \\
        V-SZZ   & 0.487 & 0.670 & 0.564 & 0.623 \\
        AG-SZZ  & 0.536 & 0.649 & 0.587 & 0.623 \\
        MA-SZZ  & 0.500 & 0.660 & 0.569 & 0.620 \\
        TSE-SZZ & 0.514 & 0.719 & 0.599 & 0.666 \\
        RA-SZZ  & 0.523 & 0.607 & 0.562 & 0.588 \\
        \midrule
        \textbf{TraceVIC@1} & {0.747} & 0.717 & {0.732} & \textbf{0.723} \\
        \textbf{TraceVIC@2} & 0.654 & 0.854 & {0.741} & \textbf{0.805} \\
        \textbf{TraceVIC@3} & 0.590 & {0.909} & {0.716} & \textbf{0.820} \\
        \bottomrule
    \end{tabular}
\end{table}

Table~\ref{tab:retrieval-results} compares TraceVIC with retrieval-based VIC identification methods. These baselines return sets of candidate commits, which generally favors recall by increasing the likelihood of retrieving the true VIC, but can reduce precision by introducing additional false positives. We therefore report both F1 and F2: F1 captures the balance between precision and recall, while F2 places greater emphasis on recall. We consider recall particularly important in this context because failing to recover the true VIC is more consequential than returning additional candidates for further manual inspection.

TraceVIC achieves a consistently stronger precision--recall trade-off across the evaluated cutoffs. At @1, it achieves the highest precision (0.747), substantially exceeding all retrieval-based baselines. Expanding to @2 increases recall from 0.717 to 0.854 while maintaining precision of 0.654, resulting in the highest F1 score (0.741) and thus the best balance between precision and recall. At @3, TraceVIC further increases recall to 0.909 and achieves the highest F2 score (0.820), while maintaining a precision of 0.590. To compare, the strongest retrieval-based baseline achieves an F1 of 0.599 and an F2 of 0.666.

These results reveal a clear trade-off across the ranking cutoffs: @1 favors precision, @2 provides the best balance between precision and recall, and @3 favors coverage, recovering more than 90\% of the ground-truth VICs. Importantly, this increased coverage is achieved with only three ranked candidates, rather than an unrestricted candidate set. Thus, TraceVIC not only improves VIC recovery but also prioritizes the most likely VICs within a small, ranked set, reducing the number of candidates that require further inspection.

\subsubsection{Selection-based Methods}
Selection-based methods reduce the candidate or revision history to a \emph{single} final VIC prediction. Their evaluation therefore corresponds to a top-1 decision: a prediction is correct if the selected commit matches a ground-truth VIC and incorrect otherwise. We compare TraceVIC against L-SZZ, R-SZZ, LLM-SZZ, AgenticSZZ, and LLM4SZZ under this formulation. L-SZZ and R-SZZ apply explicit heuristics to select one commit from the SZZ candidate set, whereas recent LLM-based approaches use model-guided reasoning over the revision history to produce a final commit prediction. Table~\ref{tab:selection_methods} summarizes these methods.


\begin{table*}[t]
\centering
\small
\caption{Overview of \textit{Selection-based} methods for VIC identification.}
\label{tab:selection_methods}
\begin{tabular}{p{1.2cm} p{7.2cm} p{3.0cm} p{3.8cm}}
\toprule
\textbf{Method} & \textbf{What it does} & \textbf{Main Distinction} & \textbf{Category} \\
\midrule

L-SZZ &
Starts from the SZZ candidate set and selects the candidate commit with the largest amount of code change. &
Largest modification &
Candidate-selection heuristic \\

R-SZZ &
Selects the most recent candidate among the commits identified by SZZ, producing a single final VIC prediction. &
Most recent candidate &
Candidate-selection heuristic \\

LLM-SZZ &
Extend V-SZZ with an LLM as the root cause line selection in the previous commit, selecting the most likely root cause line until it reaches the earliest vulnerability-inducing commit.   &
LLM semantic judgment &
LLM-based selection \\

AgenticSZZ &
Temporal Knowledge Graph per fixing commit and expanding the candidate research beyond git blame. An LLM agent navigates then navigates this graph using four tools, reasoning causally to select the single true bug-inducing commit &
Candidate space expansion using TKG traversal &
Graph-search agentic selection \\

LLM4SZZ &
Uses LLM-guided reasoning over code revisions to identify and return/rank a single final VIC prediction. &
LLM semantic judgment &
LLM-based selection \\

\bottomrule
\end{tabular}
\end{table*}

The implementations of L-SZZ and R-SZZ are obtained from the replication package of \cite{hinrichs2026back}. LLM4SZZ was executed through its own reproducibility package \cite{tang2025llm4szz}, obtained via the artifact of AgenticSZZ \cite{shi2026agenticszz}, which vendors it; we modified only its configuration layer so that repository paths, dataset locations, and the LLM endpoint are supplied through environment variables rather than hardcoded constants, leaving the prompts and algorithm unchanged. AgenticSZZ itself was run from that same artifact, and LLM-SZZ from its publicly available replication package~\cite{fan2025llm}, both under their authors' default configurations.

\begin{table}[htbp]
\small
    \setlength{\tabcolsep}{3.5pt}
    \centering
    \caption{VIC identification results comparing TraceVIC with
    \textit{selection-based} methods.}
    \label{tab:selection-results}
    \begin{tabular}{lcccc}
        \toprule
        \textbf{Method} & \textbf{Precision} & \textbf{Recall} &
        \textbf{F1} & \textbf{F2} \\
        \midrule
        L-SZZ   & 0.754 & 0.485 & 0.590 & 0.522 \\
        R-SZZ   & 0.697 & 0.448 & 0.546 & 0.483 \\
        LLM-SZZ (Mistral-7B-v0.3) 
                & 0.586 & 0.486 & 0.531 & 0.503 \\
        LLM-SZZ (Gemini 2.5 Pro)
                & 0.499 & 0.683 & 0.577 & 0.577 \\
        AgenticSZZ (Gemini 2.5 Pro)
                & 0.709 & 0.681 & 0.695 & 0.686 \\
        LLM4SZZ (Gemini 2.5 Pro)
                & 0.676 & 0.705 & 0.690 & 0.699 \\
        \midrule
        \textbf{TraceVIC@1}
                & 0.747 & {0.717} & {0.732} & \textbf{0.723}\\
        \textbf{TraceVIC@2}
                & 0.654 & 0.854 & 0.741 & \textbf{0.805}\\
        \textbf{TraceVIC@3}
                & 0.590 & 0.909 & 0.716 & \textbf{0.820} \\
        \bottomrule
    \end{tabular}
\end{table}

Table~\ref{tab:selection-results} compares TraceVIC with selection-based methods, which identify a single candidate as the predicted VIC. Because these methods produce a single prediction, TraceVIC@1 provides the most direct comparison. TraceVIC@1 achieves a precision of 0.747, recall of 0.717, F1 of 0.732, and F2 of 0.723. It therefore achieves the highest F1 and F2 among the directly comparable methods, while its precision is only slightly below L-SZZ (0.754). In contrast, L-SZZ obtains its high precision at the cost of substantially lower recall (0.485).

TraceVIC@1 also outperforms the recent LLM- and agent-based approaches. Compared with AgenticSZZ, the strongest baseline in terms of F1, TraceVIC@1 improves F1 from 0.695 to 0.732 and F2 from 0.686 to 0.723. Compared with LLM4SZZ, which achieves the highest baseline recall (0.705), TraceVIC@1 improves both precision (0.747 vs.\ 0.676) and recall (0.717 vs.\ 0.705). These results indicate that explicitly reasoning over temporal code evolution provides benefits beyond using LLM-based reasoning alone.

Although @1 provides the fairest comparison with selection-based methods, TraceVIC's ranked output additionally allows analysts to inspect a small number of alternatives. Expanding the cutoff to @2 increases recall to 0.854 and produces the highest F1 (0.741), while @3 reaches 0.909 recall and an F2 of 0.820. Thus, TraceVIC provides both strong single-candidate identification and, when higher coverage is desired, a compact ranked set that substantially increases the likelihood of recovering the true VIC. 

\subsubsection{Ranking-based Methods}
Ranking-based methods differ from the previous two categories by assigning an explicit ordering to candidate VICs. We evaluate whether the ground-truth VIC appears among the top-1, top-2, or top-3 predictions, thereby measuring both candidate recovery and prioritization.

NeuralSZZ requires special treatment because its learned ranking operates at the deletion-line level rather than the commit level. We replicated the authors' publicly released implementation~\cite{tang2023neural}. To enable a ranking-based comparison without introducing an additional commit-ranking heuristic, we preserve NeuralSZZ's learned deletion-line ordering and propagate that ordering to the commits obtained through SZZ tracing. Specifically, commits traced from higher-ranked deletion lines receive higher priority in the resulting commit ranking. We consequently evaluate two configurations, NeuralSZZ+B-SZZ and NeuralSZZ+V-SZZ, at top-1, top-2, and top-3. This evaluation preserves NeuralSZZ's learned ordering while allowing us to assess how effectively its line-level prioritization translates into VIC prioritization at the commit level.

In contrast, TraceVIC is designed to rank candidate commits directly. It first identifies the top-$k$ root-cause deletion lines, constructs temporally connected commit graphs across revisions, and then produces a ranked list of candidate VICs. We evaluate TraceVIC at top-1, top-2, and top-3. Precision@$k$ measures the proportion of the top-$k$ predictions that correspond to ground-truth VICs, while Recall@$k$ measures the proportion of ground-truth VICs recovered within the top-$k$ predictions.

\begin{table}[!t]
\small
    \centering
    \caption{VIC identification results comparing TraceVIC with
    \textit{Ranking-based} NeuralSZZ, assuming that the ranking of deletion lines reflects the ranking of their associated commits.}
    \label{tab:commit_ranking}
    \begin{tabular}{lcccc}
        \toprule
        \textbf{Method} & \textbf{Precision} & \textbf{Recall} & \textbf{F1} & \textbf{F2}\\
        \midrule
        NeuralSZZ+B-SZZ @1 & 0.724  & 0.534  & 0.615 & 0.564\\
        NeuralSZZ+B-SZZ @2 & 0.676  & 0.613 & 0.643 & 0.625\\
        NeuralSZZ+B-SZZ @3 & 0.641 & 0.636 & 0.639 & 0.637 \\
        NeuralSZZ+V-SZZ @1 & 0.713 & 0.527  & 0.606 & 0.556 \\
        NeuralSZZ+V-SZZ @2 & 0.651 & 0.597  & 0.622 & 0.607 \\
        NeuralSZZ+V-SZZ @3 & 0.619 & 0.618 & 0.619 & 0.619 \\
        \midrule
          \textbf{TraceVIC@1}
                & 0.747 & \textbf{0.717} & \textbf{0.732} & \textbf{0.723} \\
        \textbf{TraceVIC@2}
                & 0.654 & 0.854 & 0.741 & 0.805 \\
        \textbf{TraceVIC@3}
                & 0.590 & 0.909 & 0.716 & 0.820 \\
        \bottomrule
    \end{tabular}
\end{table}

As Table \ref{tab:commit_ranking} reports, compared with NeuralSZZ+B-SZZ, the strongest NeuralSZZ configuration, TraceVIC@1 improves recall from 0.534 to 0.717 and F1 from 0.615 to 0.732 while also slightly improving precision (0.747 vs. 0.724). The difference becomes more pronounced as the ranking depth increases: at @3, TraceVIC reaches 0.909 recall and 0.820 F2, compared with 0.636 and 0.637 for NeuralSZZ+B-SZZ. These results indicate that directly modeling and ranking candidate commits provides substantially stronger VIC prioritization than transferring a learned deletion-line ranking to commits through SZZ tracing.

\subsubsection{Ablation: Code-Evolution Modeling}
\label{sec:ablation}

To isolate the contribution of modeling code evolution, we compare
TraceVIC against two progressively reduced variants. \textbf{Single
Revision} restricts the model to a single program revision, removing
access to the evolution history. \textbf{w/o Temporal Edges} retains
the complete sequence of revisions and the same revision-level graph
representations and commit-ranking architecture as TraceVIC, but removes
the TEMPORAL\_FWD and TEMPORAL\_BWD edges that explicitly connect
corresponding program elements across consecutive revisions. This
ablation separates the benefit of reasoning over multiple revisions
from the additional benefit of explicitly encoding node-level
correspondences across revisions.

\begin{table}[htbp]
\small
\centering
\caption{Ablation study of code-evolution modeling. All results are
reported at top-3.}
\label{tab:temporal-ablation}
\begin{tabular}{lcccc}
    \toprule
    \textbf{Variant} & \textbf{Precision} & \textbf{Recall} &
    \textbf{F1} & \textbf{F2} \\
    \midrule
    Single Revision
        & 0.637 & 0.637 & 0.637 & 0.637 \\
    w/o Temporal Edges
        & \textbf{0.604} & 0.891 & \textbf{0.720} & 0.814 \\
    \textbf{TraceVIC}
        & 0.590 & \textbf{0.909} & 0.716 & \textbf{0.820} \\
    \bottomrule
\end{tabular}
\end{table}

Table~\ref{tab:temporal-ablation} shows that access to the evolution
history provides the largest improvement. Moving from a single revision
to multiple revision-level graphs without explicit temporal edges
increases recall from 0.637 to 0.891 and F2 from 0.637 to 0.814,
corresponding to relative improvements of 39.9\% and 27.8\%,
respectively. F1 similarly increases from 0.637 to 0.720 (13.0\%).
These results indicate that vulnerability-inducing changes are more
effectively identified when the model can reason over how the relevant
code evolves across multiple revisions rather than relying on a single
program snapshot.

Explicit cross-revision correspondences provide a smaller but
complementary benefit. Adding TEMPORAL\_FWD and TEMPORAL\_BWD edges
increases recall from 0.891 to 0.909 and F2 from 0.814 to 0.820.
This improvement comes with a modest reduction in precision, from
0.604 to 0.590, causing F1 to decrease slightly from 0.720 to 0.716.
Thus, temporal edges primarily improve coverage: by directly linking
corresponding program elements across consecutive revisions, they help
TraceVIC recover additional true VICs that are missed when revisions
are represented independently.

The ablation reveals two distinct benefits.
Reasoning over the complete revision history accounts for the majority
of TraceVIC's improvement over single-revision analysis, while explicit
node-level temporal correspondences provide an additional recall-oriented
gain. Relative to the Single Revision variant, the complete TraceVIC
model improves recall by 42.7\% and F2 by 28.7\%. These results support
our central hypothesis that reasoning over code evolution improves VIC
identification, while further showing that the benefit arises primarily
from modeling the broader evolution history rather than from any single
temporal mechanism.

\begin{tcolorbox}[colback=gray!10,colframe=gray!50,title=\textbf{Answer to RQ1},breakable]
Modeling code evolution substantially improves VIC identification. Compared with single-revision reasoning, using the full revision history improves F2 from 0.637 to 0.814, while explicit cross-revision temporal edges further increase F2 to 0.820 and recall to 0.909. TraceVIC also consistently outperforms the strongest baselines under category-appropriate comparisons: @1 for single-selection methods and @k for ranking-based methods. These results show that reasoning over the evolution history is the primary source of improvement, with explicit temporal correspondences providing an additional recall-oriented benefit.
\end{tcolorbox}

\subsection{RQ2: Generalizability}

We evaluated TraceVIC generalizability on four popular C/C++ projects: FFmpeg, ImageMagick, OpenSSL, and PHP-SRC. Our evaluation dataset is drawn from the 100 C/C++ vulnerabilities manually verified by Bao et al.~\cite{bao2022v}, excluding Linux kernel cases since our model was trained on them. From the remaining cases, one PHP-SRC instance is further excluded because its fixing commit modifies only M4 autoconf build system files containing no C source changes, yielding 79 test cases across the four projects.

\begin{table*}[htbp]
    \centering
    \small
    \caption{Generalizability results across unseen projects. 
    ``Identified CVE'' reports CVE-level success, while Precision, 
    Recall, F1, and F2 on top 3 choices are computed at the commit level over 
    all ground-truth VICs.}
    \label{tab:generalizability}
    \begin{tabular}{lcccccccc}
        \toprule
        \textbf{Project} & \textbf{Total CVE} & \textbf{Identified CVE} &
        \textbf{GT VICs} & \textbf{Correctly Identified} &
        \textbf{Precision} & \textbf{Recall} &
        \textbf{F1} & \textbf{F2} \\
        \midrule
        FFmpeg      
        & 20 & 20 & 27 & 26 & 0.867 & 0.963 & 0.912 & 0.942 \\
        
        ImageMagick 
        & 20 & 20 & 28 & 24 & 0.889 & 0.857 & 0.873 & 0.863 \\
        
        OpenSSL     
        & 20 & 20 & 44 & 37 & 0.861 & 0.841 & 0.851 & 0.845 \\
        
        PHP-SRC     
        & 19 & 18 & 36 & 26 & 0.813 & 0.722 & 0.765 & 0.739 \\
        
        \midrule
        \textbf{Overall} 
        & \textbf{79} & \textbf{78} & \textbf{135} & \textbf{113} &
        \textbf{0.856} & \textbf{0.837} &
        \textbf{0.846} & \textbf{0.841} \\
        \bottomrule
    \end{tabular}
\end{table*}

\begin{table}[htbp]
\small
\centering
\caption{Comparison with LLM4SZZ, TSE-SZZ, and NeuralSZZ (+B-SZZ) on the generalizability dataset.}
\label{tab:generalizability_comparison}

\resizebox{\columnwidth}{!}{%
\begin{tabular}{lccccc}
    \toprule
    \textbf{Method} & \textbf{Identified CVE} & \textbf{Precision} &
    \textbf{Recall} & \textbf{F1} & \textbf{F2} \\
    \midrule
    LLM4SZZ         & 60 & 0.870 & 0.444 & 0.588 & 0.493 \\
    TSE-SZZ         & 76 & 0.812 & 0.607 & 0.695 & 0.640 \\
    NeuralSZZ+B-SZZ & 75 & \textbf{0.918} & 0.566 & 0.700 & 0.613 \\
    \midrule
    \textbf{TraceVIC} & \textbf{78} & 0.856 & \textbf{0.837} &
    \textbf{0.846} & \textbf{0.841} \\
    \bottomrule
\end{tabular}%
}
\end{table}

As Table~\ref{tab:generalizability} shows, across 79 vulnerabilities from four unseen C/C++ projects, TraceVIC correctly identifies at least one valid VIC for 78 out of 79 cases, demonstrating strong transferability beyond the Linux kernel training domain.

Some vulnerabilities are associated with multiple valid inducing commits, resulting in a total of 135 ground-truth VICs across the 79 CVEs. For this reason, we report both CVE-level correctness and commit-level Precision@3, Recall@3, and F1@3. While CVE-level correctness reflects whether at least one true inducing commit is successfully identified, commit-level metrics evaluate how completely the method recovers all valid inducing commits. When multiple  VICs exist for a single vulnerability, identifying one correct commit improves CVE-level accuracy, while recall depends on recovering the full set of valid inducing commits.

Among the projects, FFmpeg achieves the strongest performance (F1@3 = 0.912) , likely because many vulnerabilities in FFmpeg involve more localized fixes and shorter commit chains, making the causal contribution easier to isolate. ImageMagick and OpenSSL also show strong and stable performance. Although OpenSSL contains the largest number of ground-truth VICs (44 for only 20 CVEs), TraceVIC still maintains high recall (0.841) and F1 (0.851), showing robustness even in projects with deeper histories and more complex vulnerability propagation.

PHP-SRC shows the lowest performance (F1=0.765), likely due to broader code propagation patterns and stronger multi-commit interactions, where vulnerabilities are distributed across larger commit histories. These project-level differences suggest that the main challenge is not the project itself, but the structural complexity of vulnerability evolution within commit histories.

Table~\ref{tab:generalizability_comparison} further evaluates TraceVIC
against the best-performing methods from the retrieval-, selection-, and
ranking-based categories on the generalizability dataset: TSE-SZZ,
LLM4SZZ, and NeuralSZZ+B-SZZ, respectively. TraceVIC correctly identifies
78 CVEs and achieves the highest recall (0.837), F1 (0.846), and F2
(0.841), while maintaining high precision (0.856).

Compared with TSE-SZZ, TraceVIC improves both precision (0.856 vs.\ 0.812)
and recall (0.837 vs.\ 0.607), resulting in a substantial improvement in
F1 (0.846 vs.\ 0.695). This comparison is particularly notable because
TSE-SZZ is retrieval-based and can return multiple candidate commits,
whereas TraceVIC explicitly ranks candidates according to their
likelihood of being the VIC.

The learning-based baselines exhibit a different precision--recall
trade-off. NeuralSZZ+B-SZZ achieves the highest precision (0.918), but
its recall is substantially lower than TraceVIC's (0.566 vs.\ 0.837).
Similarly, LLM4SZZ obtains slightly higher precision (0.870 vs.\ 0.856)
but considerably lower recall (0.444 vs.\ 0.837). Consequently, TraceVIC
achieves substantially higher F1 and F2 scores than both methods,
indicating that its improvement in VIC recovery does not come at the
cost of excessive false positives.

The gains are particularly evident for projects with deeper and more
complex revision histories, such as OpenSSL, where vulnerability-relevant
code may evolve across multiple commits. These results support the
motivation for explicitly modeling code evolution: rather than selecting
candidates according to a predefined position in the history, TraceVIC
evaluates candidate commits using their structural and temporal context.

To conclude, the results show that TraceVIC's performance extends beyond the
projects used for training and transfers effectively to the unseen
projects in the generalizability dataset. Its consistently high recall
while preserving strong precision suggests that temporal-structural
reasoning remains effective across diverse C/C++ project histories.

We further assess TraceVIC's computational cost relative to existing approaches and conduct a detailed analysis of its failure cases. Due to space constraints, these analyses are presented in Appendices~\ref{app:computational_cost} and~\ref{app:failure_analysis}, respectively.

\begin{tcolorbox}[colback=gray!10,colframe=gray!50,title=\textbf{Answer to RQ2},breakable]
TraceVIC generalizes effectively beyond its Linux-kernel training domain, identifying at least one valid VIC for 78 of 79 vulnerabilities across four unseen C/C++ projects. It achieves an overall precision of 0.856, recall of 0.837, and F1 of 0.846. Compared with the strongest baseline F1 of 0.700, TraceVIC improves F1 by 20.8\%, demonstrating that its temporal reasoning transfers effectively across previously unseen projects.
\end{tcolorbox}

\section{Related Work}

VIC identification has evolved from heuristic-based SZZ variants to
learning-based and history-aware approaches.

\subsection{SZZ Algorithms and Variants}

The original SZZ algorithm (B-SZZ)~\cite{sliwerski2005changes} applies
\texttt{git blame} to lines deleted or modified by a fixing commit and
attributes them to their most recent modifiers. Subsequent variants
primarily improve candidate filtering or history tracing. AG-SZZ~\cite{kim2006automatic},
MA-SZZ~\cite{da2016framework}, RA-SZZ~\cite{neto2018impact}, and
DJ-SZZ~\cite{williams2008szz} progressively filter non-semantic,
meta-level, refactoring, and semantics-preserving changes. R-SZZ and
L-SZZ~\cite{davies2014comparing} select candidates using recency and
change-size heuristics, respectively, while V-SZZ~\cite{bao2022v}
traces backward to the earliest reachable modification. TC-SZZ~\cite{lyu2024evaluating}
instead retains intermediate commits along the modification history.

Despite these differences, SZZ variants identify VICs through predefined
heuristics over the traced history rather than evaluating each candidate's
contribution to the vulnerable condition~\cite{hinrichs2024back}.
TraceVIC instead models this history as an evolution sequence and directly
ranks candidate commits using temporal-structural representations.

\subsection{Learning-Based SZZ Approaches}

Learning-based approaches incorporate richer structural and semantic
information. NeuralSZZ~\cite{tang2023neural} uses CodeBERT~\cite{feng2020codebert}
and graph attention over control- and data-flow relationships to rank
root-cause lines, but does not directly rank commits over their evolution
history. LLM4SZZ~\cite{tang2025llm4szz} uses LLMs to rank suspicious
statements and incorporates commit and patch context, while
LLM-SZZ~\cite{fan2025llm} combines code and natural-language context to
guide iterative tracing but ultimately selects the earliest reached
commit. These methods improve semantic reasoning over individual changes
or candidates but do not explicitly model vulnerability-relevant code
evolution across the candidate sequence for direct commit ranking.

\subsection{Temporal Reasoning over Code History}

Recent approaches incorporate software history into fault and
vulnerability analysis. FONTE~\cite{an2023fonte} propagates suspiciousness
across commit history to rank fault-inducing changes but relies on runtime
test coverage. AgenticSZZ~\cite{shi2026agenticszz} represents commit history
as a Temporal Knowledge Graph and uses an LLM agent to reason over
relationships among commits, files, functions, and developers.
CommitShield~\cite{wu2025commitshield} incorporates historical context
for vulnerability detection but classifies individual commits.

TraceVIC instead models the evolution of vulnerability-relevant code
itself by combining within-revision program structure with cross-revision
code correspondences and directly ranking candidate VICs without imposing
an earliest- or most-recent-commit rule.

\section{Threats to Validity}

\textbf{Internal Validity.}
TraceVIC traces deleted lines for deletion-bearing fixes and nearby
anchors for addition-only fixes, focusing the analysis on code associated
with the fixing region. Relevant context outside this region may therefore
be missed; however, incorporating broader context in our experiments
introduced additional false positives and reduced precision. Addition-only
fixes introduce further uncertainty because anchor selection may omit
relevant histories or introduce unrelated candidates.

\textbf{Construct Validity.}
Our evaluation primarily relies on manually verified ground-truth VICs
from Jiang et al.~\cite{jiang2024understanding}; any inaccuracies or
subjective attribution decisions are therefore inherited. Candidate
construction also depends on successful history tracing, potentially
favoring vulnerabilities with cleaner revision histories. Finally,
existing methods produce different outputs---candidate sets, single
predictions, or rankings. We therefore evaluate these formulations
separately, although differences in output semantics may still affect
cross-method comparisons.

\textbf{External Validity.}
Our primary evaluation uses Linux kernel vulnerabilities, with
generalizability evaluated on four unseen C/C++ projects. Thus, although
the evaluation covers multiple repositories, the findings may not
generalize to other languages or ecosystems. Differences in coding
practices, commit granularity, and vulnerability-report quality may also
affect performance.

\section{Conclusion and Future Work}

We reformulate VIC identification as reasoning over code evolution rather
than positional selection and introduce \textbf{TraceVIC}, a
temporal-structural framework that models vulnerability-relevant code
across revisions. Our ablation study shows that multi-revision reasoning
provides the primary improvement over single-revision analysis, with
explicit cross-revision correspondences providing additional gains.
TraceVIC outperforms retrieval-, selection-, and ranking-based methods
and generalizes to unseen C/C++ projects. Future work will address
vulnerabilities induced by interacting commits, complex code changes
such as large refactorings, and languages beyond C/C++.


\cleardoublepage
\appendix


\bibliographystyle{plainurl}
\bibliography{main}

\appendix
\input{appendix}

\end{document}

%% file: appendix.tex
\section*{Ethical Considerations}

This study analyzes publicly available source code, commit histories,
and previously disclosed vulnerabilities. It does not involve human
participants or interaction with deployed systems. The principal
ethical consideration is the potential dual use of vulnerability
history analysis. We mitigate this risk by evaluating only previously
disclosed vulnerabilities and not releasing undisclosed vulnerability
or exploit information.

\section*{Open Science}
All artifacts required to evaluate the core contributions of this 
paper including the dataset, source code, pre-trained models, 
evaluation scripts, configuration files, and documentation are 
made available via an anonymous repository at \url{https://anonymous.4open.science/r/TraceVIC/}.
\begin{itemize}

    \item \textbf{Source Code.} Full implementation of the end-to-end 
    VIC identification pipeline, including CPG construction, the Graph 
    Attention Network encoder, and CommitTransformer.

    \item \textbf{Datasets.} All benchmark datasets in preprocessed form, 
    sourced from publicly available repositories, are available.

    \item \textbf{Pre-trained Models.} Model checkpoints for all 
    configurations evaluated in the paper, enabling direct reproduction 
    of reported results without retraining.

    \item \textbf{Evaluation Scripts.} Scripts reproducing all quantitative 
    results, ablation studies, and baseline comparisons, each annotated 
    with expected inputs, outputs, and runtime.

    \item \textbf{Configuration Files.} All hyperparameter settings, 
    model architecture configurations, and training parameters in 
    structured format (YAML/JSON).

\end{itemize}
The repository includes complete instructions to reproduce all 
experimental results reported in the paper.

\section{Dataset Filtering}
\label{app:dataset_filtering}

The original dataset contains 1,349 CVE instances. We applied filtering criteria at each stage of the TraceVIC pipeline, excluding instances that are fundamentally incompatible with our analysis. Table~\ref{tab:dataset_filtering} summarizes the filtering process, which reduces the dataset from 1,349 to 780 CVE instances.

Prior to any processing, we excluded 89 instances where the relevant source files exceed 5,000 lines of code. In these cases, the vulnerability-inducing commit corresponds to the initial commit in which all code was first introduced into the file, making commit-level differentiation impossible regardless of the method applied.

At the graph construction stage, we excluded 15 instances involving assembly files, as the graph construction pipeline is designed for C source code and does not extend to assembly-level analysis. We additionally excluded 25 instances where the files referenced in the fixing commit are absent from the repository, preventing graph construction entirely.

At the V-SZZ traversal stage, we excluded 440 instances where the fixing commit SZZ traversal did not produce any candidate vulnerability-inducing commit. Without a candidate commit chain, commit ranking cannot be performed.

\begin{table}[htbp]
\centering
\caption{Dataset Filtering Summary}
\label{tab:dataset_filtering}
\begin{tabular}{lc}
\toprule
\textbf{Filtering Criterion} & \textbf{Instances Excluded} \\
\midrule
Source files exceeding 5,000 lines   & 89  \\
Assembly files                       & 15  \\
Missing repository files             & 25  \\
No candidate commit   & 440 \\
\midrule
\textbf{Original Dataset}            & \textbf{1,349} \\
\textbf{Final Dataset}               & \textbf{780} \\
\bottomrule
\end{tabular}
\end{table}

\section{CPG Construction}
\label{app:cpg}

\subsubsection*{Compilation Context Reconstruction}

C programs depend heavily on preprocessing, header inclusion, and compiler-specific configurations, making accurate analysis contingent on reconstructing the full compilation context. Without this context, header dependencies remain unresolved, leading to incorrect or incomplete ASTs. TraceVIC addresses these challenges by explicitly reconstructing compilation environments using Bear~\cite{Bear}, which intercepts the build process and records compilation commands, including header paths, preprocessor definitions, and compiler flags, producing a \texttt{compile\_commands.json} database.

\subsubsection*{Semantic Unit Extraction}

With the compilation database in place, TraceVIC uses Clang~\cite{lattner2004llvm} to parse each C source file into an AST, from which it extracts semantic units that serve as graph nodes. These units include function declarations, control structures, and expressions, providing a structured representation at the level of granularity required for vulnerability analysis. We adopt AST-level granularity (statements and expressions) as it provides a balance between semantic expressiveness and computational tractability. Because the full compilation context has been reconstructed, the Clang--Bear integration resolves C-specific constructs such as struct definitions, pointer operations, macro expansions, and preprocessor directives, so the extracted representation faithfully captures the true semantics of the code.

\subsubsection*{Control and Data Flow Extraction}

Semantic nodes alone are insufficient to reason about vulnerability behavior, as vulnerabilities arise not from isolated code elements but from how they interact. TraceVIC uses Joern~\cite{yamaguchi2014modeling} to extract Control Flow Graph (CFG) and Data Flow Graph (DFG) edges from the parsed C code, encoding execution order and variable dependencies respectively. Joern's default representation operates at the token level; TraceVIC discards Joern's node representation and retains only the extracted flow edges, which are then integrated with the higher-level semantic units derived from Clang.

\subsubsection*{Semantic-Flow Fusion and Refinement}

TraceVIC integrates Clang nodes and Joern edges into a unified CPG by aligning both representations at the source line level. For each flow edge identified by Joern, the corresponding Clang AST units are located and a directed edge is introduced between them. The unified graph is then filtered using the fixing commit's diff: only nodes corresponding to lines present in the diff are retained, focusing the representation on vulnerability-relevant code changes.

\section{Node Correspondence Matching}
\label{app:node_correspondence}

For each history entry $h_t = (\text{line\_num}_t, \text{code}_t)$ produced by V-SZZ for a given revision $c_t$, TraceVIC identifies the corresponding node $v \in V_t$ in the CPG of that revision using a three-level priority cascade.

\textbf{Level 1 (line range and code prefix).} A node $v$ is selected if $h_t$'s line number falls within $v$'s line range and a prefix of length $\lambda$ of $h_t$'s code appears in $v$'s code, or vice versa:
\begin{equation}
\begin{split}
\text{lineBeg}(v) \leq \text{line\_num}_t \leq \text{lineEnd}(v) \\
\text{and } \bigl(\text{code}_t[:\lambda] \subseteq v.\text{code} \lor v.\text{code}[:\lambda] \subseteq \text{code}_t\bigr)
\end{split}
\end{equation}

\textbf{Level 2 (line range only).} If no Level~1 match is found, TraceVIC selects any node $v$ satisfying $\text{lineBeg}(v) \leq \text{line\_num}_t \leq \text{lineEnd}(v)$. This handles cases where minor code reformatting prevents exact prefix matching.

\textbf{Level 3 (code prefix only).} If neither Level~1 nor Level~2 yields a match, TraceVIC selects any node $v$ satisfying the code prefix condition above. This handles cases where line numbers shift due to insertions or deletions in adjacent code.

If none of the three levels produces a match for a given revision, no temporal edge is added for that revision pair. When a revision's CPG is empty, TraceVIC inserts a synthetic node with the line number and code content from the history entry to preserve chain connectivity. We use $\lambda = 20$ characters throughout all experiments.

\section{Evaluation Metrics}
\label{app:metrics}

\textbf{Root Cause Line Identification.} We report Precision, Recall, and F1-Score at ranks 1, 2, and 3. Precision@$k$ measures the proportion of the top-$k$ ranked lines that are true root cause lines. Recall@$k$ measures the proportion of true root cause lines successfully identified within the top-$k$ ranked lines. F1@$k$ is the harmonic mean of Precision@$k$ and Recall@$k$.

\begin{equation}
P@k = \frac{1}{N} \sum_{i=1}^{N} \frac{|\{\text{rank}_1^{(i)}, \ldots, \text{rank}_k^{(i)}\} \cap GT_i|}{k}
\end{equation}

\begin{equation}
R@k = \frac{1}{N} \sum_{i=1}^{N} \frac{|\{\text{rank}_1^{(i)}, \ldots, \text{rank}_k^{(i)}\} \cap GT_i|}{|GT_i|}
\end{equation}

\begin{equation}
F1@k = 2 \cdot \frac{P@k \cdot R@k}{P@k + R@k}
\end{equation}

where $N$ is the total number of test cases, $GT_i$ is the set of ground truth root cause lines for test case $i$, and $\text{rank}_j^{(i)}$ is the $j$-th ranked line for test case $i$.

\section{Graph Encoder Equations}
\label{app:encoder_equations}

This appendix provides the full equations for the edge-aware graph-attention layer described in Section~\ref{sec:encoder}.

\textbf{Input projections.} For each node $i$ at head $h$:
\begin{align}
Q_i^h &= h_i W_Q^h, \quad K_i^h = h_i W_K^h, \quad V_i^h = h_i W_V^h
\end{align}
where $W_Q^h, W_K^h, W_V^h \in \mathbb{R}^{d \times d_h}$ and $d_h = d / H$.

\textbf{Edge-type-biased attention.} For each directed edge $(j \to i)$:
\begin{equation}
e_{ij}^h = (Q_i^h \cdot K_j^h) \cdot d_h^{-\frac{1}{2}} + b^h_{t_{ij}}
\end{equation}
where $t_{ij} \in \{0,\ldots,6\}$ is the edge type index and $b^h_{t_{ij}}$ is a learned scalar drawn from an embedding table $\mathbf{B} \in \mathbb{R}^{7 \times H}$.

\textbf{Normalization and aggregation.}
\begin{align}
\alpha_{ij}^h &= \frac{\exp(e_{ij}^h)}{\sum_{k \in \mathcal{N}(i)} \exp(e_{ik}^h)} \\
z_i^h &= \sum_{j \in \mathcal{N}(i)} \alpha_{ij}^h V_j^h
\end{align}

\textbf{Output projection and residual.}
\begin{align}
h_i' &= \text{LayerNorm}\Bigl(h_i + \text{Dropout}\bigl(W^O (z_i^1 \| \cdots \| z_i^H)\bigr)\Bigr) \\
h_i'' &= \text{LayerNorm}\Bigl(h_i' + \text{Dropout}\bigl(W_2\,\text{GELU}(W_1 h_i')\bigr)\Bigr)
\end{align}
where $W_1, W_2 \in \mathbb{R}^{d \times d}$. TraceVIC applies $L = 2$ layers in all experiments.

\section{Commit Attribution Equations}
\label{app:phase2_equations}

This appendix provides the full equations for the commit attribution stage described in Section~\ref{sec:attribution}.

\subsection*{Input Projection}

Node embeddings from the frozen encoder ($\mathbb{R}^{768}$) are projected to the internal dimension $d'$:
\begin{equation}
\tilde{h}_v = \text{Dropout}(\text{GELU}(\text{LayerNorm}(W_{\text{in}} h_v'')))
\end{equation}
where $W_{\text{in}} \in \mathbb{R}^{768 \times d'}$.

\subsection*{Correspondence-Aware Attention Pooling}

Keys and values are projected for all nodes:
\begin{equation}
K_v^h = \tilde{h}_v W_K^h \in \mathbb{R}^{d_h'}, \quad V_v^h = \tilde{h}_v W_V^h \in \mathbb{R}^{d_h'}
\end{equation}
where $d_h' = d'/H$. Each node is assigned a query based on its role:
\begin{equation}
q_v^h = \begin{cases} q_{\text{corr}}^h & \text{if } v \text{ is a TEMPORAL\_FWD target} \\ q_{\text{local}}^h & \text{otherwise} \end{cases}
\end{equation}
where $q_{\text{corr}}^h, q_{\text{local}}^h \in \mathbb{R}^{d_h'}$ are learned parameters. Attention logits and weights, normalized over all nodes in commit $c_t$:
\begin{align}
e_v^h &= (q_v^h \cdot K_v^h) \cdot d_h'^{-\frac{1}{2}} \\
\alpha_v^h &= \frac{\exp(e_v^h)}{\sum_{u \in \mathcal{S}(c_t)} \exp(e_u^h)}, \quad v \in \mathcal{S}(c_t)
\end{align}
Per-head commit embeddings are aggregated and projected:
\begin{align}
\mathbf{c}_t^h &= \sum_{v \in \mathcal{S}(c_t)} \alpha_v^h V_v^h \\
\mathbf{c}_t &= \text{LayerNorm}\bigl(W_O (\mathbf{c}_t^1 \| \cdots \| \mathbf{c}_t^H)\bigr) \in \mathbb{R}^{d'}
\end{align}

\subsection*{Commit-Level Transformer}

Self-attention over the commit sequence:
\begin{align}
e_{ts} &= \frac{(\mathbf{c}_t W_Q)(\mathbf{c}_s W_K)^T}{\sqrt{d_h'}}, \quad \alpha_{ts} = \frac{\exp(e_{ts})}{\sum_{r} \exp(e_{tr})} \\
\mathbf{c}_t' &= \text{LayerNorm}\Bigl(\mathbf{c}_t + \text{Dropout}\Bigl(W_O \sum_{s} \alpha_{ts}\,\mathbf{c}_s W_V\Bigr)\Bigr) \\
\mathbf{c}_t'' &= \text{LayerNorm}\bigl(\mathbf{c}_t' + \text{Dropout}(\text{FFN}(\mathbf{c}_t'))\bigr)
\end{align}
where $\text{FFN}(x) = W_2\,\text{GELU}(W_1 x)$ with hidden dimension $4d'$.

\subsection*{Ranking Head}

\begin{equation}
s_t = W_2 \cdot \text{Dropout}(\text{GELU}(W_1\, \mathbf{c}_t''))
\end{equation}
where $W_1 \in \mathbb{R}^{d' \times d'/2}$ and $W_2 \in \mathbb{R}^{d'/2 \times 1}$.

\subsection*{Loss Function}

\textbf{Soft targets.} For $C$ candidate commits with ground-truth set $G$ and smoothing factor $\epsilon$:
\begin{equation}
\tilde{y}_t = \begin{cases} \dfrac{1 - \epsilon}{|G|} & t \in G \\[6pt] \dfrac{\epsilon}{C - |G|} & t \notin G \end{cases}, \qquad \tilde{y}_t \leftarrow \frac{\tilde{y}_t}{\sum_j \tilde{y}_j}
\end{equation}

\textbf{Predicted probabilities.}
\begin{equation}
p_t = \frac{\exp(s_t / \tau)}{\sum_{j=1}^{C} \exp(s_j / \tau)}, \qquad \bar{p}_{\text{gt}} = \frac{1}{|G|} \sum_{t \in G} p_t
\end{equation}

\textbf{Focal cross-entropy loss.}
\begin{equation}
\mathcal{L}_{\text{focal}} = \alpha \cdot (1 - \bar{p}_{\text{gt}})^{\gamma} \cdot \Bigl(-\sum_{t=1}^{C} \tilde{y}_t \log(p_t + \epsilon)\Bigr)
\end{equation}

\textbf{Margin loss.} Let $N = \{1,\ldots,C\} \setminus G$:
\begin{equation}
\mathcal{L}_{\text{margin}} = \frac{1}{|G| \cdot |N|} \sum_{i \in G} \sum_{j \in N} \max(0,\; \delta - (s_i - s_j))
\end{equation}

\textbf{Combined loss.}
\begin{equation}
\mathcal{L} = \mathcal{L}_{\text{focal}} + \lambda\, \mathcal{L}_{\text{margin}}
\end{equation}

TraceVIC uses $\tau = 0.1$, $\epsilon = 0.05$, $\gamma = 2.0$, $\alpha = 1.0$, $\delta = 0.5$, and $\lambda = 0.5$ in all experiments.

\section{Addition Cases Structural Pattern}
\label{app:addition_cases_structural}

Manual analysis of addition-only CVEs turned up one recurring shape:

\begin{center}

\begin{tabular}{@{}ll@{}}

\toprule

\textsc{Role} & \textsc{Status in pre-fix file} \\

\midrule

Anchor operation    & present \\

Missing companion   & \emph{absent}, added by the fix \\

Downstream operation & present \\

\bottomrule

\end{tabular}

\end{center}

\noindent The vulnerability is the absence of the companion statement (a bounds check, a resource release, a lock guard), not an error in the anchor. The VIC is the commit that introduced the anchor operation without its required companion, so blaming the anchor line, rather than a deleted line that does not exist, traces directly to it.

\paragraph{Extraction procedure.}

For each contiguous block of added lines, we extract up to three anchor lines from the post-fix file: lines already present in the pre-fix version that sit next to, or enclose, the insertion point.

\begin{itemize}[nosep,leftmargin=*]

\item \textbf{Above}: the nearest non-noise line preceding the block, for \emph{missing-cleanup} defects (the fix adds a release for a resource acquired above).

\item \textbf{Below}: the nearest non-noise line following the block, for \emph{missing-guard} defects (the fix adds a check for an operation present below).

\item \textbf{Signature}: the enclosing function signature, located by scanning upward with brace-depth tracking. A function signature is rarely modified after the VIC introduces it, so it works as a stable fallback when neighbouring lines are noise or post-date the VIC.

\end{itemize}

\noindent A noise filter strips structural boilerplate (\texttt{goto}, \texttt{return -ERRNO}, labels, braces, \texttt{break}/\texttt{continue}) so anchors land on meaningful code rather than scaffolding. Anchors are deduplicated, and mapped from post-fix to pre-fix line coordinates. Each is emitted as a synthetic deletion line in $\mathcal{D}'(V)$, so the candidate-chain construction (Eq.~1--2) and everything downstream run without modification.

\begin{figure}[t]
  \centering
  \includegraphics[width=\columnwidth]{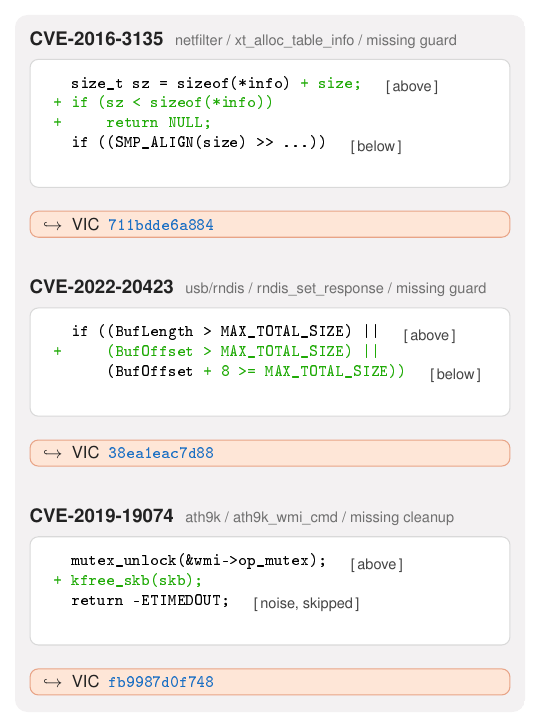}
  \caption{Anchor-line extraction for three addition-only VFCs.
  TraceVIC identifies nearby pre-existing code as anchors for
  tracing the vulnerability-relevant revision history.}
  \label{fig:anchor-examples}
\end{figure}

Figure~\ref{fig:anchor-examples} illustrates anchor extraction for three addition-only fixes and shows why multiple anchor types are considered: depending on the structure of the fix, different anchors may provide access to the vulnerability-relevant history.

\textbf{\noindent\emph{CVE-2016-3135} (netfilter, \texttt{xt\_alloc\_table\_info})}.
The fix inserts an integer-overflow guard. The \textsc{above} anchor, the overflowing computation \texttt{sz = sizeof(*info) + size}, traces to the ground-truth VIC (\texttt{711bdde6a884}); the \textsc{below} anchor does not. This is the case for extracting more than one anchor type: relying on just one would have missed it here.

\textbf{\noindent\emph{CVE-2022-20423} (USB RNDIS, \texttt{rndis\_set\_response})}.
The fix adds a missing bounds-check clause to an existing compound \texttt{if}. Both anchors trace to \texttt{38ea1eac7d88}. The fixing commit's own message reads \texttt{Fixes:~38ea1eac7d88}, a kernel-maintainer attribution independent of our dataset, and it matches our prediction exactly.

\textbf{\noindent\emph{CVE-2019-19074} (ath9k, \texttt{ath9k\_wmi\_cmd})}.
The fix inserts a single \texttt{kfree\_skb(skb)} to close a memory leak. The \textsc{above} anchor traces to \texttt{fb9987d0f748}, the ground-truth VIC. The nearest \textsc{below} candidate, \texttt{return~-ETIMEDOUT}, is dropped by the noise filter: without filtering, error-handling boilerplate like this would dilute the candidate pool. The two surviving anchors agree, leaving a pool of exactly one commit.

\section{Computational Cost}
\label{app:computational_cost}

Accuracy is only part of the practical picture: a method that must be re-run whenever new vulnerabilities are disclosed is limited as much by its cost per query as by its precision. We therefore measured wall-clock time for TraceVIC and the two strongest LLM-based baselines over the same 755 Linux kernel cases. TraceVIC was run on a single NVIDIA RTX 4090 GPU; the LLM baselines were served by Gemini 2.5 Pro under their authors' default configurations, so their timings reflect API latency rather than local computation.

\begin{figure}[htbp]
    \centering
    \includegraphics[
    width=\columnwidth,
]{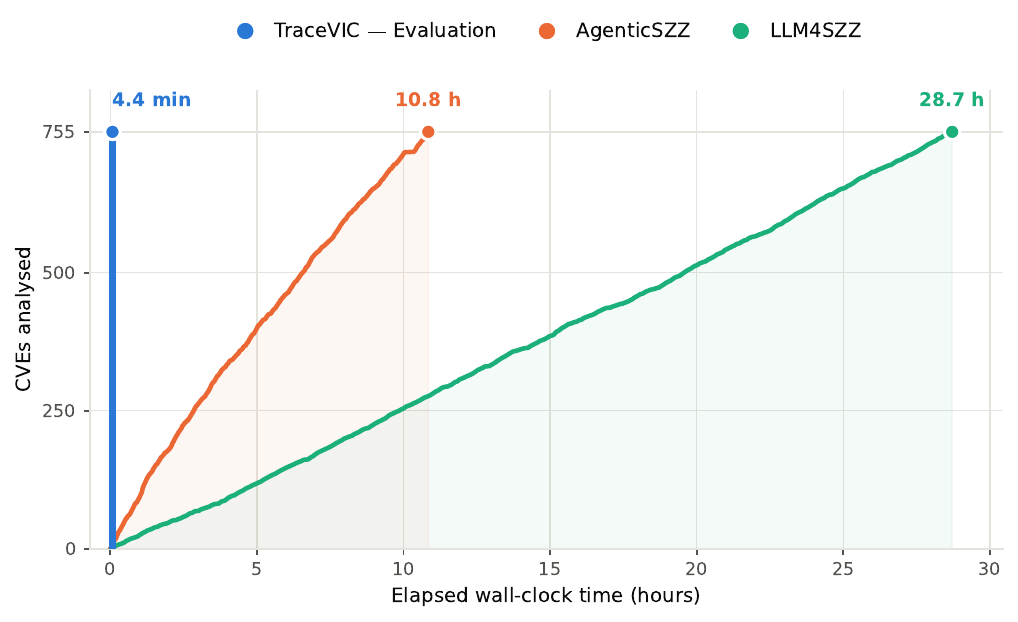}

    \caption{Cumulative CVEs analyzed against elapsed wall-clock time over the 755-case
Linux kernel dataset.}
    \label{fig:fig_runtime_race}
\end{figure}

Figure~\ref{fig:fig_runtime_race} reports cumulative CVEs analyzed against elapsed time. TraceVIC ranks all 755 cases in 4.4 minutes—0.35 seconds per case—compared with 10.8 hours for AgenticSZZ and 28.7 hours for LLM4SZZ, speedups of roughly 147× and 391× respectively. The gap arises from where the analysis cost is incurred. The LLM-based methods issue repeated model queries per case: AgenticSZZ traverses its temporal knowledge graph through successive agent tool calls, and LLM4SZZ performs iterative context-enhanced assessment over candidate commits. Both incur a network-bound, per-case cost that recurs on every invocation and scales with the length of the candidate commit chain. TraceVIC instead front-loads its cost: graph construction and node encoding are performed once per repository revision and cached, after which ranking a vulnerability requires a single forward pass over the cached temporal subgraphs. 

\section{Failure Analysis}
\label{app:failure_analysis}

We manually analyzed 21 failure cases where TraceVIC failed to correctly identify the VICs in the Linux kernel dataset to better understand the main sources of error. The most common failure pattern was commits with a large number of deleted lines (7 cases, 1.1\%), followed by large refactoring performed together with the fix (6 cases, 0.9\%). Vulnerability propagation through later refactoring accounted for 3 cases (0.4\%), while code movement without actual code changes and downstream extensions each appeared in 2 cases (0.3\%). Overall, most failures occur when strong surrounding structural changes overshadow the true vulnerability signal.

\begin{itemize}[leftmargin=*]

\item \textbf{Large number of deleted lines:} 
VFCs containing many deleted lines make root cause identification more difficult. When 16 or more lines are deleted, the large number of non-root-cause deletions introduces substantial noise, diluting the signal of the true root cause and making it harder for the model to isolate the semantically critical line.

\item \textbf{Large refactoring with the fix:} 
When developers fix a vulnerability while also performing major code cleanup or refactoring, the structural changes produce stronger representational signals than the actual vulnerability fix. As a result, the model may prioritize refactoring deletions over the true root cause.

\item \textbf{Code movement without code change:} 
In some cases, the bug is fixed only by reordering lines rather than changing code content. Larger co-moved blocks create stronger signals than the isolated root cause line, causing the model to rank them higher.

\item \textbf{Downstream extensions masking the origin:} 
Later commits may extend the vulnerable design by adding more related code around it. These commits create stronger lexical and structural overlap with the fixing commit than the true VIC, leading to incorrect attribution.

\item \textbf{Vulnerability propagated by later refactoring:} 
When a later refactoring preserves the vulnerable line unchanged but restructures the surrounding function, the stronger contextual similarity with the fixing commit can displace the true origin commit in the ranking.

\end{itemize}

These results show that TraceVIC is most challenged when the true vulnerability signal is hidden by larger structural changes, especially large refactorings and noisy deletion-heavy fixes. Future work should focus on improving robustness to such cases by incorporating finer-grained semantic change analysis that can better separate true causal changes from surrounding refactoring noise.